\documentclass[pre,a4paper,twocolumn,superscriptaddress]{revtex4}

\usepackage{graphicx}
\usepackage{esdiff}
\usepackage{color}
\usepackage{amsmath}

\newcommand{\HH}{Hodgkin-Huxley}

\begin{document}

\title{How is information transmitted in a nerve?}

\author{Michel Peyrard}
\email[E-mail: ]{Michel.Peyrard@ens-lyon.fr}
\affiliation{Laboratoire de Physique de l'Ecole Normale Sup\'{e}rieure de Lyon,
  46 All\'{e}e d'Italie, 69364 Lyon C\'{e}dex 07, France}


  \date{\today}


\begin{abstract}
In the last fifteen years a debate emerged about the validity of the famous
Hodgkin-Huxley model for nerve impulse. Mechanical models have been
proposed. This note reviews the experimental properties of the nerve impulse
and discusses the proposed alternatives. The experimental data, which rule out
some of the alternative suggestions, show that, while the Hodgkin-Huxley model
may not be complete, it nevertheless includes essential features that should
not be overlooked in the attempts made to improve, or supersede, it.

\smallskip
{\em Revised version, accepted for publication in the Journal of Biological
  Physics.} 
\end{abstract}

\maketitle

\section{Introduction}

In high-school biology courses a standard experiment shows how a small
voltage applied to a dead-frog muscle can induce its
contraction. Actually it reproduces the first observation made by Luigi
Galvani in the $18^{\mathrm{th}}$ century. In 1850 Hermann von Helmholtz designed
an experiment to measure the velocity of the signal that propagates along the
sciatic nerve of a frog \cite{HELMHOLTZ1850,HELMHOLTZ-CRAS}.
A quantitative description of the
propagation of an electrical signal in a nerve was proposed in 1952 by
A.L.\ Hodgkin and A.L.\ Huxley \cite{HODGKIN1952} after a careful series of
experimental investigations. For a long time this Hodgkin-Huxley model,
recognised in 1963 by the Nobel Prize for Physiology or Medicine,
stayed as the unquestioned basic model of this phenomenon, which
launched a new field of research \cite{VANDENBERG,SCHWIENING}.
According to this
model the nerve impulse is due to voltage-controlled flows of sodium and
potassium ions through the axon membrane.

\smallskip
However the phenomena are more complex. Experiments also detect heat transfer
and a slight deformation of the axon together with the electrical signal. In
the last fifteen years, this lead some scientists to raise questions about the
\HH\ model and even to propose an alternative picture in which the propagation
of a mechanical signal is the main feature
\cite{HEIMBURG,HOLLAND}. 

\smallskip
This article reviews the main experimental data accumulated over more than a
century on the propagation of the nerve impulse, paying attention to some
aspects which are particularly relevant for the ongoing discussion on the
validity of the \HH\ model. Then it discusses the proposals
to replace, or complete, this model. The final discussion contains some
comments on these attempts.

\section{Experimental studies of the electrical properties of axons}

The German school of ``organic physicists'' played a major role in creating
modern physiology in the second half of the $19^{\mathrm{th}}$ century
\cite{SEYFARTH}. Hermann von Helmoltz (1821--1894),
who worked in Heidelberg, was the first to
measure the velocity of the signal along a nerve, and
Julius Bernstein (1839 -- 1917), who was
trained under von Helmholtz, designed a clever apparatus which allowed him to
record the shape of the nerve impulse \cite{BERNSTEIN1868}. Then he related
the potential difference across the membrane to the Nernst theory in a paper
which founded the theoretical analysis of the phenomenon
\cite{BERNSTEIN1902}. However, in his studies Bernstein focused his attention
on the negative part of the pulse, which is associated to a potassium flux. It
is Ernest Overton (1865 -- 1933), working in W\"{u}rzburg, who pointed out the
essential role of sodium for the excitation of a muscle \cite{OVERTON}.

\smallskip
Following this earlier period, the most significant results came from England,
with the work of Alan Lloyd Hodgkin (1914 -- 1998)
(Trinity College, Cambridge) and his
co-workers, particularly Andrew Fielding Huxley (1917 -- 2012),
and from the USA where Kenneth Stewart Cole (1900 -- 1984)
and Howard James Curtis (1906 -- 1972) (Columbia
University, New York) developed powerful experimental methods. After a first
visit there in 1939, Hodgkin kept collaborating with this group, which
significantly influenced his research. These results are important in the
present context of the discussion of the \HH\ model because they,
unambiguously, demonstrated that a flow of ions across the axon membrane was
determining the shape of the voltage pulse which carries information along a
nerve. The course of the work, which is a mixture of careful planning and
accidental observations has been vividly reported in a conference of Hodgkin
in 1976 \cite{HODGKIN1976}.

\subsection{Evidence for electrical transmission in nerve (Hodgkin-1937)}

The goal of this early study \cite{HODGKIN1937} was to determine whether a
local excitation was able to excite a neighbouring region, i.e.\ to get an
evidence that a signal was actually propagating along the nerve. Hodgkin
created a blocking region by freezing the axon over 3 to 5 mm, or by pressing
it between two blocks of ebonite. Although the electrical pulse could not pass
through, he noticed that some signal was nevertheless leaking through the
blocking domain, and was making the nerve highly excitable on the other
side. Following the arrival of a pulse on one side of the blocked region,
on the other side a new signal could be generated by a much smaller electrical
excitation than in an unperturbed region of the axon. Although Hodgkin did not
discuss any mechanical response of the axon, in the context of the present
discussions on the nature of the signal, this example shows that
freezing the mechanical displacements in some region of the axon does not block
the transmission of some signal. This pleads against a purely mechanical
driving mechanism for the nerve impulse.

\subsection{Searching for the mechanism of the nerve impulse}
\label{sec:searching}

Hodgkin alone or with the help of different collaborators in Cambridge, as
well as Curtis and Cole in the USA, carried an impressive series of systematic
experiments which provided the basis for the development of the \HH\
model. The results of these measurements should not be overlooked in any
theory of the nerve impulse.

\smallskip
In an attempt to detect local effects of the membrane excitation,
Hodgkin was able to isolate a single unmyelinated nerve fibre of a crab 
(with a great piece of luck as he says himself \cite{HODGKIN1976}) and this
allowed him to detect sub-threshold potentials, i.e.\ signals which are not
sufficient to launch a nerve impulse but nevertheless locally modify the
properties of the membrane \cite{HODGKIN1938}. This was the first evidence of
a local response in a nerve fibre.

A decisive step was made by Cole and Curtis \cite{COLE1939}, inspired by the
``membrane theory'' of Bernstein \cite{BERNSTEIN1902} who postulated a
decrease of the membrane resistance during a pulse. They managed to measure
the impedance of the membrane of the giant axon of a squid during the passage
of a pulse created several centimetres away, using a highly sensitive
electrical bridge. The choice of the squid axon was an element of their
success because it has a diameter which can reach $0.5\;$mm and segments
$3-8\;$cm long could be prepared. The other important element was to perform
the measurements at low temperature ($4-8^{\circ}$C) which reduced the
conduction velocity and enhanced the signal captured by the bridge.

The measurements only detected a small change of the membrane capacity but a
big drop of its resistance when the voltage pulse passed. The change of the
capacity depends on the frequency but never excess 10\%, with an average of
2\%. In the rest state the membrane of the axon is non-conducting, but, during
the passage of a pulse, Cole and Curtis found that its resistance dropped by a
factor of about 36. The time of rise of the conductance could not be precisely
determined but it was estimated to be less than $100\;\mu$s.

\smallskip
The experiment of Cole and Curtis did not investigate the role of specific
ions. This was further studied in a series of experiments which varied the
ionic content of the medium around the nerve fibres.

Hodgkin \cite{HODGKIN1939} showed that the velocity of the pulse decreases by
about 30\% when the outside medium changes from sea water to oil, which is an
important indication for the models.

Curtis and Cole \cite{CURTIS1942} made
measurements with one electrode inside the axon and another outside.
For the first time it gave a precise view of the membrane potential. The
resting potential ($ V_{\mathrm{inside}} - V_{\mathrm{outside}} $ in \HH\
notation) is about
$-50\;$mV, while the action potential is positive and differs by about
$110\;$mV from the resting potential.  Replacing
the $Na^+$ ions of the outside medium by $K^+$ causes a dramatic drop of the
action potential. The values of the resting and action potentials were later
confirmed by Hodgkin and Huxley \cite{HODGKIN1945}, but, as recognised by
Huxley himself \cite{HUXLEY2002} the explanations that they proposed for the
sign reversal within the pulse were wrong. It is only soon after that they
began to consider an increase in membrane permeability specific for sodium
ions, which was confirmed by a study by Hodgkin and Katz
\cite{HODGKIN-KATZ}. All these experiments are very challenging because the
electrodes can become polarised during the measurement, and thus alter the
potentials to be measured. Hodgkin and Katz  developed specific electrodes
that they could introduce inside the axon and their paper describes several
tests that they made to validate their results. Using various external
solutions, they demonstrated the crucial role of the sodium, and the sharp
rise of membrane permeability to sodium when the action potential arrives. In
agreement with Overton's observations \cite{OVERTON} they could show that
lithium ions also show an effect very similar to sodium although, on the
long term, lithium damages the axon. Therefore all data started to fit together
nicely to set the stage of the \HH\ model proposed in 1952 \cite{HODGKIN1952}.

\smallskip
This picture, inferred from experiments immersing axons in various solutions,
was later confirmed by a direct observation of the flow of ions through the
membrane in a series of papers by Hodgkin and Keynes in 1955 using radioactive
tracers \cite{HODGKIN1955a,HODGKIN1955b}. The experiments showed that, during
the nerve impulse, both $Na^+$ and $K^+$ move down concentration gradients,
i.e.\ their transport is {\em passive}, contrary to the slower transport which
brings the axon back to its rest state in the recovery process. It uses
metabolic energy, it is highly temperature dependent, and can be inhibited by
dinitrophenol contrary to the passive transport during the pulse. In these
voltage clamp experiment, which can impose a fixed potential difference
even if the concentration of the ions is modified, Hodgkin and Keynes managed
to show that the potassium flux is {\em not proportional to concentration} but
increases more steeply \cite{HODGKIN1955b}. This convinced them that the ions
do not move independently from each other, and they showed that all their
observations could be well reproduced by a model in which the $K^+$ ions move
along a chain of potassium selective sites which stretch through the membrane,
and that all $n$ sites in each chain are occupied by a potassium.
It is quite remarkable that a careful analysis of macroscopic experiments
managed to determine the features of the ion channels which were detected only
years later.

\bigskip
Another study by Hodgkin and Katz \cite{HODGKIN-KATZ-b} is also very important
in the context of current discussions on the basic mechanism of the nerve
impulse. It is their investigation of
the effect of temperature on the electrical
activity of the giant axon of the squid from $-1^{\circ}$C to
$40^{\circ}$C. The resting potential was practically constant up to
$20^{\circ}$C and dropped at higher temperature. The action potential showed a
gradual evolution, with a slight decrease in amplitude up to $20^{\circ}$C and
then a faster drop above $30^{\circ}$C. Over the whole temperature range, the
change in the time scale of the pulse is gradual, but very significant. At
lower temperature the nerve pulse becomes very broad, and moreover the rise
and fall times have different temperature dependencies. The fall time grows
much more than the rise time when temperature decreases. This lead Hodgkin and
Katz to suggest different mechanisms for these two processes, which is
consistent with the current knowledge that they involve different ion
channels.

\section{The nerve impulse is not only an electrical signal}
\label{sec:nonelec}

Although a large part of the efforts to understand the transmission of signals
along nerves focused on the electrical aspects, the phenomena are more
complex as shown by calorimetric and mechanical measurements.

\subsection{Thermal effects}
\label{sec:thermal}

In 1848 Helmholtz failed to detect any heat effect associated to the nerve
impulse, and the data remained controversial for several decades
\cite{FENG1936}. It is only more than eighty years later that reliable
evidences of a very weak heat effect associated to the nerve impulse could be
obtained but repeated stimulation was necessary so that the relative timing
between the heat release or absorption and the electrical pulse could not be
determined \cite{FENG1936}. A very small resting heat production could also be
detected by putting a frog nerve in a nitrogen atmosphere. Depriving the nerve
from oxygen appears to stop some oxidative processes, leading to a slow
decline of the resting heat production.

\smallskip
More modern measurements managed to follow the details of the heat exchanges
for a single impulse in non-myelinated nerves
\cite{HOWARTH1968,HOWARTH1975}. The pulse causes first an emission of heat,
and, in a second stage an absorption which almost compensates the
emission. The measurements show a gradual evolution between $4^{\circ}$C and
$15^{\circ}$C. The magnitude of the positive heat decreases and the interval
between the stimulus and the negative heat increases. Replacing sodium by
lithium does not change the overall picture but the heat emission is reduced
by about 20\% while the absorption increases by a similar amount. In the whole
temperature range, there is a close agreement between the heat emission and
the rising phase of the the action potential, and between the absorption and
the falling phase of the potential.

\smallskip
It is tempting to connect the heat effects to the energy needed to charge or
discharge a capacitance, but the quantitative analysis shows that this
simple ``condenser model'' does not account for all the observed heat
exchanges. The authors of these measurements speculated that a great part of
the heat exchanges could come from changes in the entropy of the nerve
membrane when it is depolarised and repolarised.

\medskip
However the ``condenser model'' is clearly oversimplified. For a membrane in a
conducting fluid the charge distribution is not only located at the
surface. Very recently a new thermodynamics analysis has been carried out
\cite{DE-LICHTERVELDE}. A full electrostatic model of the charged bilayer has
been established. Assuming that the equilibration of the diffuse layer is
sufficiently fast compared to the dynamics of the action potential, the paper
uses a Poisson-Boltzmann approach to derive the charge distribution, and then
compute the electrostatic energy. The entropy associated to the electric field
takes into account the polarisation of the water dipoles in the diffuse
layers which reduces entropy, as well as the entropy changes inside the lipid
membrane which can be deduced from the temperature dependence of the membrane
capacitance. The results support the idea that the heat exchanges measured
when the nerve impulse propagates have an electrostatic origin.
The results heavily depend on the membrane surface charge and only a
calorimetric measurement performed together with the recording of the
transmembrane potential with an electrode inside the axon might fully confirm
the electrostatic origin of the heat of nervous conduction, but, as discussed
in detail in \cite{DE-LICHTERVELDE}, this looks highly plausible.

\subsection{Mechanical and structural changes}
\label{sec:mechanical}

Early measurements showed signs that the excitation potential is accompanied
by some mechanical and structural changes in the axon \cite{TASAKI1968}. Small
changes in turbidity and birefringence were observed. Axons immersed in
anilinonaphtalene sulfonic acid, the fluorescence of which is extremely
sensitive to conformational changes of various macromolecules, also showed
fluorecence changes associated to nerve excitation \cite{TASAKI1968}.

\smallskip
Laser interferometry managed to detect rapid changes in the diameter of an
axon, which take place when an action potential progresses along the giant
axon of the crayfish \cite{HILL-BC}. The recorded deformation starts
$250\;\mu$s {\em after} the excitation by the pulse and starts by a contraction
which peaks $400\;\mu$s later. Then the diameter returns to normal before
showing a slow expansion to finally recover its initial size after about
$4\;$ms. The simultaneous recording of the electrical and
mechanical pulses shows that, besides
the delay of the mechanical deformation, one also observes that
this deformation occurs on a
significantly longer time scale. The overall displacement is very small,
ranging from $3\;$\AA\ to $25\;$\AA\ in different nerve preparations. It
decreases when the nerve deteriorates. These experiments could also show that
the deformation is directly linked to the action potential. If the electrical
stimulation of the axon is reduced to 90\% of the value that triggers a pulse,
the mechanical deformation of the axon is not observed.

\smallskip
This result was later confirmed by optical measurements using the near field
at the end of an optical fibre brought in close proximity of the axon
and by piezo-electric measurements of the pressure at the axon surface
\cite{IWASA}. The mechanical displacement reaches $50$ to $100\;$\AA\ for crab
nerves \cite{IWASA2}.
Clues on the mechanism of the swelling of axons were provided by
studying volume transitions observed in synthetic and natural ionic gels
by varying the ratio of monovalent and bivalent ions
\cite{TASAKI1992}. Changing the concentration in $Na^+$, $Li^+$, $K^+$ ions of
a medium containing gel beads showed sharp transitions of the bead diameters,
associated to thermal effects, which could be understood as structural
transitions in the gel. Similar observations were made with the squid giant
axon \cite{TASAKI1999a,TASAKI1999b},
suggesting that the change in ionic concentrations
around the membrane, induced by the nerve impulse, could lead to transitions in
the membrane structure responsible for the mechanical deformation associated
to the pulse. Heat exchanges, which accompanies these transitions could
contribute to the thermal effects recorded when a pulse passes.

\section{The debate on the \HH\ model}

\subsection{The \HH\ model}

In the famous article which introduced their model \cite{HODGKIN1952}, Hodgkin
and Huxley explain that it is built upon a series of measurements of the
flow of electric currents through the surface membrane of the squid giant
axon. From their experiments, they deduced that the main features of the nerve
impulse result from a transient increase of the sodium conductance of the
membrane, which leads to a strong flow of $Na^+$ ions towards the inside of
the axon generated by the gradient of sodium concentration across the
membrane. This step, leading to a rise of the action potential $V =
V_{\mathrm{inside}} - V_{\mathrm{outside}}$, is followed by a slower but
maintained increase in potassium conductance. In the rest state the potassium
concentration is greater inside the axon than outside, so that potassium tends
to flow out of the axon, causing a decrease of the action potential, which,
after a small overshoot, finally comes back to its resting value. However
Hodgkin and Huxley went well beyond this qualitative picture. From their
measurements they managed to propose a set of differential equations which
model the sodium and potassium conductances. These equations could not be
solved analytically but Hodgkin and Huxley relied on a pioneering approach to
obtain solutions that they could quantitatively compare with their
experiments. They performed what was probably the first numerical
simulation in biological physics.
However, as the the EDSAC (Electronic delay storage automatic
calculator), built in the Cambridge Mathematical Laboratory, which
they later used for their studies, was
not immediately available, they had to carry lengthy calculations on a manual
calculator \cite{SCHWIENING}.
Their article \cite{HODGKIN1952} contains a
detailed section on the numerical methods, which explains for instance how an
iterative scheme could be used to determine one unknown parameter, the
propagation speed of the impulse. This approach allowed a thorough test of the
model, not only for the pulse but also for subthreshold responses.

\smallskip
The model is clearly only focused 
on the {\em electrical aspects} of signal propagation along nerves. Hodgkin
and Huxley took a great care to discuss the limitations of their model, but
did not discuss other physical phenomena, such as thermal effects.
although they were certainly aware of their existence. Presumably they
considered the action potential to be the dominant phenomenon. Moreover the
model has not been established from the basic principle of physics and
chemistry, which would have naturally introduced other phenomena, for instance
through a thermodynamics analysis. In the discussion of the paper the authors
wrote that ``the agreement [with experiments] must not be taken as evidence
that our equations are anything more than an empirical description of the time
course of the changes in permeability to sodium and potassium''. Therefore it
is not surprising that the model can be discussed and completed. However, if
the model stood out as the main model to describe the nerve impulse for about
70 years, it is because Hodgkin and Huxley based
their conclusions on a large set
of detailed experiments that they thoroughly analysed. This is probably why
the model, established before the knowledge of the existence and structure of
ionic channels could propose equations for the variation of the membrane
conduction which turned out to match structural data of the ionic channels
that were discovered much later. For instance, for potassium channels
Hodgkin and Huxley noticed that, for the opening, the conductance versus time
needed a third or fourth-order equation to be fitted, while the closing could
be described by a first order equation. This lead them to a model in which
the potassium channel opening is controlled by 4 sites which should
simultaneously occupy a certain position in the membrane. Molecular dynamics
simulations of a voltage gated potassium channel, carried 60 years later
\cite{JENSEN}, are perfectly consistent with this view because they confirm
that 4 voltage sensitive domains must be up before the pore can reopen after
closure. Thus, although Hodgkin and Huxley took great care in stressing that
``the interpretation given is unlikely to provide a correct picture of the
membrane'', they managed to extract a lot from their voltage clamp
measurements, which strengthen the credibility of the model that they
proposed. The structure and basic function of sodium and potassium
ionic channels were discovered about 30 years after the proposal of Hodgkin
and Huxley \cite{CATERALL2000}, and this confirmed their insight.

\smallskip
Of course the \HH\ model is not perfect, however the precise analysis of the
voltage clamp techniques that underlie its equations lead to a wide acceptance
of the hypothesis of channels mediating ionic flow across the membrane
and the \HH\ model kinetics describes most of the classic experiments
fairly well although some experiments showed that the picture may be
oversimplified. In spite of its successes, as noticed in Sec.~\ref{sec:nonelec},
the model does not describe all phenomena associated to nerve signalling, so
that some scientists noticed that ``given the many
experimental features not explained within the \HH\ theory, it is surprising
that it remains an unchallenged dogma'' \cite{HEIMBURG}.

\subsection{A mechanical model for the propagation of the nerve impulse}
\label{sec:mechmodel}

And indeed the dogma has been challenged, in particular with proposals that
the dominant effect in nerve signalling might be mechanical rather than
electrical \cite{HEIMBURG}. T. Heimburg and A. Jackson suggest that the
nerve impulse could actually propagate as a localised deformation of the
axon \cite{HEIMBURG}.
In most of the physical systems a localised deformation, which, in
Fourier space, is an infinite combination of signals of different wavelengths,
tends to
spread as it propagates due to dispersion because the different wavelengths
propagate at different speeds. However, in some systems, the effect of
dispersion can be compensated by nonlinearity, leading to solitary waves,
which may have particle-like properties, which gave them their name of
``solitons'' \cite{MPsolitons}. Heimburg and Jackson
noticed that lipid membranes generally
display order-disorder phase transitions in a temperature range which is not
far from physiological temperatures. Heating can destroy the lateral order
of the molecules, which absorbs heat and leads to a swelling of the
structure. This structural change modifies the volume-
and area-compressibility of the membrane.
As a result, in the vicinity of the transition,
the speed of sound depends on $\rho_A$, its mass
per unit area. Therefore the equation for
the propagation of a mechanical disturbance $\Delta \rho_A$ contains a
nonlinear contribution associated to the variation of the lateral
compressibility near the transition. And because the thermal exchanges
occurring at the transition are slow processes, the speed of sound, which is
also a function of the specific heat as shown by the thermodynamics Maxwell
relations, depends on the frequency (and wavelength) of the mechanical
disturbance, so that the equation for the propagation of the mechanical
disturbance also includes dispersion. Heimburg and Jackson show that the
signs of the contributions
are such that nonlinearity can compensate dispersion. Using
standard expansions, they derive an equation for $\Delta \rho_A$ which has
some similarities with the Boussinesq equation, a standard equation in soliton
theory \cite{MPsolitons}.

\smallskip
This analysis suggests that, in the vicinity of the order-disorder transition
of lipid membranes, a mechanical perturbation can therefore propagate as a
quasi-soliton. Owing to the exceptional properties of solitons, in particular
their ability to move by preserving their shapes in the presence of
perturbations, or even in collisions with other solitons, its is tempting
to conclude that the main phenomenon that lies behind the propagation of the
nerve impulse is the compensation between nonlinearity and dispersion
which allows
the motion of narrow mechanical disturbances. In this view the electrical signal
studied by Hodgkin and Huxley is not the dominant mechanism but a secondary
effect that would be slaved to the mechanical disturbance. The deformation of
the membrane could affect the proteins that form the ion channels, and
therefore induce the ionic flow through the membrane.

\smallskip
The idea of solitons propagating along lipid membranes is interesting and it
would deserve studies to confirm it in some experiments. This is probably why
it sounded sufficiently attractive to appear as an alternative to the \HH\
model of nerve impulse. It is supported by the experiments that recorded a
variation of the diameter of the axon in the region of the pulse
\cite{HILL-BC,IWASA,IWASA2}. The heat exchange at the transition also
appeared as a candidate to explain the discrepancy between the measured
thermal exchanges that accompany the nerve impulse and the evaluations deduced
from the condenser theory \cite{HOWARTH1968,HOWARTH1975}.
Moreover an experiment which showed
that action potential launched towards each other in some axons which allow
orthodromic (normal) and antidromic (inverse) propagation can pass through
each other \cite{GONZALEZ-PEREZ2014}, in analogy with a remarkable property of
solitons, sounds as a strong argument supporting the mechanical soliton
picture for the nerve impulse.

\bigskip
However, while it might play a role in cell mechanics, the
theory proposed by Heimburg and Jackson does not stand up to close scrutiny
regarding the propagation of nerve impulse when it is
confronted to experiments.

\smallskip
\textit{(i)} The link to a phase transition in the axon membrane, occurring at
a particular temperature $T_c$ close to physiological temperature, is a
serious constraint. First, while such a transition has been observed in
unilamellar vesicles, bovine lung surfactant, and two bacterial membranes, it
has not been reported for the axon membrane \cite{HEIMBURG}. But the main
problem is that it gives a specific role to a temperature $T_c$ while
experimental investigations of the effect of temperature on the nerve impulse,
from $-1^{\circ}$C to $40^{\circ}$C \cite{HODGKIN-KATZ-b} do not show any
discontinuity or qualitative change around a specific temperature, but rather
a gradual evolution (Sec.~\ref{sec:searching}).

\smallskip
\textit{(ii)} The argument of a disagreement between the measured thermal
effects and the evaluations of the condenser theory, that could be explained
by the thermal exchanges associated to the latent heat of the transition is
not very strong because the condenser theory is oversimplified and more
careful evaluations of the energy exchange associated to ion transfers
\cite{DE-LICHTERVELDE} do not conclude to such a disagreement, although the
conclusion on this point may still be open.

\smallskip
\textit{(iii)} Similarly the conclusion drawn from the possibility of nerve
pulses to pass through each other should be taken with caution because other
experiments, using different axons, contradict this result
\cite{TASAKI1949}. Moreover the \HH\ model does not preclude such a crossing
of the pulses for some values of its parameters \cite{ASLANIDI}, which could
explain why some axons show the survival of colliding nerve pulses while
others do not.

\smallskip
\textit{(iv)} In \cite{HEIMBURG} the mechanical properties of the axon
membrane are those of a lipid layer, but actually the membrane is much more
complex. As shown in \cite{XU-2013}, actin, spectrin and associated proteins
form a cytoskeletal structure in axons. This must make the membrane much more
rigid than a simple lipid bilayer, drastically reducing nonlinear effects in
its mechanical distortions.

\smallskip
\textit{(v)} The strongest argument against the theory of Heimburg and
Jackson \cite{HEIMBURG} is actually provided by the measurements of the
deformation of the axon \cite{HILL-BC}, which they present as supporting their
idea (Sec.~\ref{sec:mechanical}).
As they used an interferometric method and simultaneously record the
electrical signal, Hill et al.\ could determine the precise timing of the two
phenomena. The mechanical signals begins by a contraction which starts
$250\;\mu$s {\em after} the arrival of the electrical pulse, and therefore it
cannot be the mechanical signal that causes the electrical signal.
Owing to the time scale of the electrical nerve impulse, the delay of
$250\;\mu$s is really significant. Moreover
the time tracks displayed in Fig.~3 of \cite{HILL-BC} show that the mechanical
signal lasts significantly longer than the electrical pulse.
Rather than supporting the idea that a mechanical could induce the electrical
pulse, as proposed in \cite{HEIMBURG}, the data of \cite{HILL-BC} support
instead the proposal that Tasaki \cite{TASAKI1992} presented after a series of
measurements \cite{IWASA,IWASA2,TASAKI1999a,TASAKI1999b}. The swelling of the
axon might be related to a structural transition, but, instead of a
transition induced by a temperature change, it would be a transition caused by
the change of the ionic environment around the membrane, which {\em follows}
the electrical pulse.

\medskip
Thus, although the idea of a soliton-like propagation of the nerve impulse
might look attractive, a precise confrontation with experimental facts cannot
support this proposal.

\bigskip
Another recent view of the nerve impulse \cite{KOTTHAUS} considers another type
of soliton, belonging to the class of envelope solitons \cite{MPsolitons} in
which a carrier wave is modulated by a localised envelope function to generate
a localised wavepacket. In this model the carrier wave would be an oscillation
of the dipolar orientation of the water molecules in the vicinity of the
membrane. The dipolar interactions would lead to forces applied to the
membrane, generating a coupling between a mechanical disturbance and the
dipolar reorientations so that the dipolar signal would ``surf on the
capillary waves propagating along the axon''. This picture appears to suffer
from several weaknesses:

\smallskip
\textit{(i)} The picture is only qualitative and no quantitative evaluation
has been made to explain how such a combined dipolar-mechanical signal could
stay localised. Capillary waves are dispersive and would tend to spread. A
quantitative mechanism has to be presented to show, in a convincing way, that
the coupling tends to maintain the necessary localisation, and that it has the
strength to do it.

\smallskip
\textit{(ii)} The dissipation in the mechanical signal would presumably be
very high. At the macroscale, capillary waves can propagate with a rather
small dissipation. But, as shown by Purcell in a a beautiful article ``Life at
low Reynolds number'' \cite{PURCELL}, at the microscopic scale phenomena are
very different, and the viscosity of water plays a much stronger role.
Moreover, considering the frequencies of the order of 100 kHz considered as
plausible in \cite{KOTTHAUS} for the capillary waves, at the
scale of the axon disturbances, dissipation would damp out
the motion very quickly due to water viscosity but also the losses within the
membrane itself.

\smallskip
\textit{(iii)} But the main objection to the scheme proposed in \cite{KOTTHAUS}
is that it assumes that, once the signal is launched by some ion transfers,
the action potential moves on without charge currents. This contradicts the
observations of local ionic currents through the membrane \cite{NEHER} and
the evidence of the ionic flows using radioactive tracers
\cite{HODGKIN1955a,HODGKIN1955b} (Sect.~\ref{sec:searching}).

\subsection{Completing the \HH\ model}

While the picture of a nerve pulse dominated by a mechanical signal does not
stand up in front of a critical examination, it does not mean that the \HH\
model cannot be completed to include some phenomena that were deliberately
left out by Hodgkin and Huxley, who focused their attention
on the electrical phenomena
only. Several attempts have been made around this idea, and they probably
point out to a direction that could improve our understanding of nerve
signalling.

\bigskip
A.\ El Hady and B.B.\ Machta \cite{EL-HADY} studied the mechanical surface
waves which accompany the propagation of the action potential. Their viewpoint
is completely different from those of Heimburg \cite{HEIMBURG} and Kotthaus
\cite{KOTTHAUS}. They don't consider that the mechanical signal is the main
signalling pathway. Instead they {\em assume} that the axon carries an
electrical pulse, defining the action potential, without making any hypothesis
on the origin of this pulse, which could be the \HH\ pulse, or have a
different origin. This pulse {\em drives} the membrane deformation and they
compute the response of the axon to this driving. In this approach the axon is
viewed as an elastic and dielectric tube filled by a viscous fluid. The
elastic energy is stored in the deformation of the tube, and the kinetic
energy is carried by the motion of the axoplasmic and extracellular viscous
fluid, which moves according to the Navier-Stokes equation. The displacement
of the membrane is expressed as a {\em linear} function of the forces due to
the action potential. Therefore this model does not consider any mechanical
nonlinearity, which is probably legitimate if the membrane is actually
strengthened by a cytoskeleton, as observed in \cite{XU-2013}. However this
approach does not require nonlinearity to localise the mechanical distortion
because the shape of the signal is imposed by that of the action potential,
which is spatially localised. Besides the energy exchanges due to charge
transfer through the membrane, this model predicts an additional heat effect
due to the isothermal distortion of the membrane because its free energy
depends on its area, which is locally modified. Using parameters estimated
from what we know of the axon, the calculation leads to results in the range
of the experimental observations. And, in contrast to the model of
\cite{HEIMBURG}, as the mechanical distortion is a consequence of the
electrical pulse, it is natural that it follows the action potential arrival
with a small delay as observed in \cite{HILL-BC}.

\bigskip
The approach of Engelbrecht et al. \cite{ENGELBRECHT} takes a similar
viewpoint that the electrical signals are the carriers of information in
nerves and trigger all other processes, but it is more ambitious because it
tries to describe the coupling between the mechanical aspects (fluid flows and
membrane deformation) and the action potential, instead of assuming that the
mechanical component is slaved to the electrical one. In their view the
channels in the membrane can be open and closed under the influence of the
electrical signals but also by mechanical inputs. The shape of the action
potential is therefore not assumed, but instead described by a simplified
version of the \HH\ model, the FitzHug-Nagumo model, initially proposed by
FitzHugh as a model for the axon \cite{FITZHUGH-1961} and then built and
further studied by Nagumo et al. \cite{NAGUMO} as a transmission line using
tunnel diodes, for possible applications in electronics and signal
processing. The mechanical signal is described by an equation which includes
nonlinearity and dispersion, in the spirit of \cite{HEIMBURG}, and a
phenomenological coupling between the electrical and mechanical components is
added. This coupling is expressed as a function of the {\em
    variation} of the fields rather than their instantanous values, which puts
  emphasis on dynamical effects.

\smallskip
The motivation of this approach is interesting because it should provide some
understanding of the mechanisms that link the electrical and mechanical
signals. However the experimental data on this coupling, which appears to be
too complex to be described from first principles of physics and chemistry, are
still sparse, and little is known on the effect of mechanical constraints to
control the ion channels. As a result the assumptions made by Engelbrecht et
al. are  difficult to validate so that, in its present stage, this approach
does not actually bring a further understanding of nerve signalling.

\smallskip
Finding the actual mechanisms behind the coupling between the electrical and
mechanical components is a challenge to reach a meaningful extension of the
\HH\ model. An interesting suggestion has been made by Krichen and Sharma
\cite{KRICHEN}. Piezoelectricity is a well known coupling between mechanical
strain and electrical polarization, but, for uniform strains it only applies
to materials which lack a mirror symmetry. However, as pointed out by Krichen
and 
Sharma, if the strain itself does not have a mirror symmetry, even a
centro-symmetric medium such as a fluid or a membrane can exhibit an
electromechanical coupling. This is flexoelectricity. Moreover, as membranes
are highly flexible, they can show very large strain gradients. Even if the
coupling coefficient is small, the resulting effect can be large. Chen et. al.
\cite{CHEN-Haoyu} used this idea to propose an axon model which couples the
distortion of the axon and the action potential. It is a two-way coupling
because a strain gradient can induce an electrical polarization, and the
action potential can cause a local distortion. Their approach can treat
myelinated and unmyelinated axons. This is an attractive idea which would
deserve further experimental and theoretical studies to be fully validated.

\section{Discussion}

In spite of alternatives introduced in the last 15 years, the answer to the
question raised in the title ``How is information transmitted in a nerve ?''
still appears to be ``as an electrical pulse'', approximately described by the 
\HH\ model proposed in 1952. Of course the answer that it provides is
oversimplified. This model was developed mostly from data recorded on the
giant squid axon, and it should be amended to describe other axons, but it is
nevertheless very likely to describe the essence of the phenomena involved in
nerve signalling.

From the start the model was not designed to be complete as Hodgkin and
Huxley only investigated the electrical component of the signal. Experiments
have shown that, as for most biological phenomena, nerve signalling is complex,
also involving a local deformation of the axon and thermal exchanges. This
lead some authors to challenge the main ideas of the \HH\ model and even to
put forward mechanisms in which the electrical signal is not dominant, such as
a mechanical soliton as the carrier of information. However, as shown in
Sec.~\ref{sec:mechmodel}, this idea cannot stand in front of a careful
examination of the experimental facts. In the case of a phenomenon as complex
as the nerve impulse, a first-principle model is still out of reach and one
must rely on phenomenological models. This can nevertheless be fruitful if
this approach is built on a detailed analysis of the experimental
observations, without neglecting some data which, at a first glance, may seem
non-essential. This is exactly what Hodgkin and Huxley did. They established
their model after years of experiments and thoughts, which allowed them
to predict phenomena or properties which had not yet been observed, such as
some features of the ion channels. Instead the models assuming a dominant
mechanical signal, although they were motivated by the observed change of the
diameter of the axon that accompanies the action potential, neglect some
elements (Sec.~\ref{sec:mechmodel}).
The deformation starts after the rise of the electrical pulse so
that causality excludes that it could generate it.

The mechanisms of anesthesia have been discussed as a possible test of the
models for the propagation of the nerve impulse. One could wonder whether
their study could help decide between the two alternatives, electrical or
mechanical. This is questionable because anesthetics probably exert their
action on synaptic transmission rather than axonal conduction
\cite{FRANKS1994}. Nevertheless there are many evidences that anesthetics act
on ion channels \cite{YAKAMURA},
which is a hint that nerve signals are electrical rather than
mechanical. Various studies have shown that anesthetics act directly on
proteins rather than on lipids \cite{FRANKS1994,EL-DIN}. There are however
cases in which the membrane is involved, as demonstrated recently for inhaled
anesthetics \cite{PAVEL}, but the actual target is nevertheless an ion channel
and the membrane lipids only play an intermediate role.

Thermal effects have also been suggested as a mean to solve
the dilemma between the electrical and mechanical views of nerve
signaling. This has two aspects, first whether the pulse is adiabatic, i.e.\
exothermic and endothermic contributions are equal, and second what is the
magnitude of thermal effects. Adiabaticity cannot make the difference
between the two alternatives. The
view that the Hodgkin-Huxley action potential must be dissipative because it
involves currents through resistors is oversimplified. Experiments
\cite{HOWARTH1968,HOWARTH1975} show that exothermic effects are followed by an
endothermic process of about the same magnitude. In the \HH\ model this is
understandable because during the rise of the action potential the $Na^+$ ions
move down the potential gradient giving rise to heat release, while, in
the second stage the $K^+$ ions move up the potential gradient but down their
concentration gradient, converting heat into capacitive energy
\cite{EL-HADY}. In the soliton picture, thermal effects come from the latent
heat of a reversible phase transition in the membrane, so that the overall
thermal effect vanishes. The second aspect, the magnitude of the thermal
effect, could, in principle, lead to a conclusion because the thermal effect
due to a phase transition in the membrane should be significantly greater than
the magnitude measured by experiments, ruling out the soliton model. However
the measurements are very difficult and, for the fastest pulses, they may not
have a sufficient temporal resolution to catch the full magnitude of the
thermal exchanges \cite{HEIMBURG2020}. Nevertheless, as experiments observe
that replacing sodium by lithium modifies the magnitude of the thermal effect,
and as theories that go beyond the simple condenser model conclude that a
proper thermodynamic analysis of the processes involved in the Hodgkin-Huxley
model is compatible with the experimental observations, the studies of the
thermal effects appear to favor the electrical view. As pointed out in
\cite{EL-HADY}, an additional contribution to thermal effects could
nevertheless come from the stretching of the membrane which accompanies the
action potential in recent models. Its order of magnitude is well below the
contribution of latent heat in the soliton picture.

\medskip

This does not mean that the proposals challenging the \HH\ model have been
useless. They stimulated further thoughts and lead to models that
combine electrical and mechanical effects. The goal of a model is not to
reproduce experimental facts but to show what are the underlying phenomena
which lead to these observations and allow further developments. Including
mechanical distortion in a nerve-impulse model
makes sense if it actually contributes to the process,
which would be the case if ion channels are not only sensitive to voltage but
also to forces or membrane distortions, as suggested by some studies
\cite{BEYDER}. However, establishing a reliable model of the axon coupling
electrical and mechanical signals will certainly need further
experimental investigations at the scale of ion channels.

A model may also be useful to understand additional phenomena.
For nerve signalling,
the role of the myeline layer deserves attention. Within the \HH\ model,
or its simplified version the FitzHugh-Nagumo model, it is easy to show how an
extra layer reducing the capacitance of the membrane can speed-up the signal,
but the role of myeline also introduces constraints on the ionic flow so that
its effect is not straightforward to predict \cite{FITZHUGH-1962}.

The \HH\ model has many parameters, and they could vary from cell to cell or
with changes in the external medium. A promising line of investigations is to
try to reduce the parameter space by looking how some parameter combinations
may be enough to determine the main features of the model \cite{ORI2018}, and
what is the stability of the results when the parameters or environment
conditions change. Recent investigations \cite{ORI2018} suggest that viewing
the \HH\ model in a low-dimensional space may bring a deeper understanding of
cell excitability.

The validity of a model also depends on the scale of interest. The \HH\ model
smoothes out the effect of individual ion channels in its continuous
equations. When the noise due to individual channels becomes relevant, discrete
stochastic models may be more appropriate \cite{STRASSBERG}. On the other hand
the \HH\ model has also been challenged for studies at the scale of a full
neural network \cite{MEUNIER}. Therefore this model is certainly not the
ultimate model for nerve signalling. It can be completed to include additional
physical phenomena or modified for special purposes, but it does not deserve
to be discarded simply because it does not describe all the complexities of
the nerve impulse. 

\section*{Conflict of interest}
The author declares that he has no conflict of interest.


\begin{thebibliography}{xx}

\bibitem{HELMHOLTZ1850}
H. Von Helmholtz:
{  Messungen über den zeitlichen Verlauf der
Zuckung animalischer Muskeln und die Fortpflanzungsgeschwindigkeit der
Reizung in den Nerven. }
Archiv f\"ur Anatomie, Physiologie und wissenschaftliche
Medicin, 276-364 (1850)

\bibitem{HELMHOLTZ-CRAS}
H. Von Helmholtz:
{  Note sur la vitesse de propagation de l'agent nerveux dans les nerfs
  rachidiens.}
Compte Rendus de l'Académie des Sciences (Paris) {\bf XXX}, 204-206 (1850)
{  Deuxi\`eme note sur la vitesse de propagation de l'agent nerveux}
Compte Rendus de l'Académie des Sciences (Paris) {\bf XXXIII}, 262-265 (1851)
(available at:
https://www.academie-sciences.fr/archivage\_site/activite/hds%
/textes\linebreak[0]/tsf\_Debru1.pdf

\bibitem{HODGKIN1952}
A.L. Hodgkin and A.F. Huxley:
{  A quantitative description of membrane current and its application to
  conduction and excitation in nerve.}
J. Physiol. {\bf 117}, 500-544 (1952)

\bibitem{VANDENBERG}
J.I. Vandenberg and S.G. Waxman:
{  Hodgkin and Huxley and the basis for electrical signalling:
  a remarkable legacy still going strong.}
J. Physiol. {\bf 590.11} 2569-2570 (2012)

\bibitem{SCHWIENING}
C.J. Schwiening:
{  A brief historical perspective: Hodgkin and Huxley.}
J. Physiol. {\bf 590.11}, 2571-2575 (2012)

\bibitem{HEIMBURG}
T. Heimburg and A.D. Jackson:
{  On soliton propagation in biomembranes and nerves.}
PNAS {\bf 102} 9790-9795 (2005)

\bibitem{HOLLAND}
L. Holland, H.W. de Regt and B. Drukarch:
{  Thinking About the Nerve Impulse: The Prospects for the Development
  of a Comprehensive Account of Nerve Impulse Propagation.}
Frontiers in Cellular Neuroscience {\bf 13}, art. 208 (2019)

\bibitem{SEYFARTH}
E.-A. Seyfarth:
{  Julius Bernstein (1839 -- 1917): pioneer neurobiologist
  and biophysicist.}
Biological Cybernetics {\bf 94}, 2-8 (2006)

\bibitem{BERNSTEIN1868}
J. Bernstein:
{  Ueber den zeitlichen Verlauf der negativen Schwankung des
  Nervenstroms.}
Pfl\"{u}gers Archiv {\bf 1}, 173-207 (1868)

\bibitem{BERNSTEIN1902}
J. Bernstein:
{  Untersuchungen zur Thermodynamik der bioelectrischen Str\"{o}me.}
Pfl\"{u}gers Archiv {\bf 92}, 521-562 (1902)

\bibitem{OVERTON}
  E. Overton:
{  Beitr\"age zur allgemeinen Muskel- und Nervenphysiologie. II Ueber die
  Unentbehrlichkeit von Natrium- (oder Lithium-)Ionen für den Contractionsact
  des Muskels.}
Pfl\"ugers {\bf 92} 346-386 (1902)

\bibitem{HODGKIN1976}
A.L. Hodgkin:
{  Chance and design in electrophysiology: an informal account of certain
  experiments on nerve carried out between 1934 and 1952.}
J Physiology {\bf 263} 1-21 (1976)

\bibitem{HODGKIN1937}
A.L. Hodgkin:
{  Evidence for electrical transmission in nerve. Part I.}
J. of Physiology {\bf 90} 183-210 (1937)

\bibitem{HODGKIN1938}
A.L. Hodgkin:
{  The subthreshold potentials in a crustacean nerve fibre.}
Proc. Roy. Soc. London {\bf B 126}, 87-121 (1938)

\bibitem{COLE1939}
K.S. Cole and H.J. Curtis:
{  Electric impedance of the squid giant axon during activity.}
J. of General Phisiology {\bf 22}, 649-670 (1939)

\bibitem{HODGKIN1939}
A.L. Hodgkin:
{  The relation between conduction velocity and the electrical resistance
  outside a nerve fibre.}
J. of Physiology {\bf 94} 560-570 (1939)

\bibitem{CURTIS1942}
H.J. Curtis and K.S. Cole:
{  Membrane resting and action potential from the squid giant axon.}
J. of Cellular and Comparative Physiology {\bf 19}, 135-144 (1942)

\bibitem{HODGKIN1945}
A.L. Hodgkin and A.F. Huxley:
{  Resting and Action potentials in single nerve fibres.}
J. Physiol. {\bf 104}, 176-195 (1945)

\bibitem{HUXLEY2002}
A.F. Huxley:
{  Hodgkin and the action potential.}
J. of Physiology {\bf 538} 2 (2002)

\bibitem{HODGKIN-KATZ}
A.L. Hodgkin and B. Katz:
{  The effect of sodium ions on the electrical activity of the giant axon of
  the squid.}
J. of Physiology {\bf 108} 37-77 (1949)

\bibitem{HODGKIN1955a}
A.L. Hodgkin and R.D. Keynes:
{  Active transport of cations in giant axons from Sepia and Loligo.}
J. of Physiology {\bf 128} 28-60 (1955)

\bibitem{HODGKIN1955b}
A.L. Hodgkin and R.D. Keynes:
{  The potassium permeability of a giant nerve fibre.}
J. of Physiology {\bf 128} 61-88 (1955)

\bibitem{HODGKIN-KATZ-b}
A.L. Hodgkin and B. Katz:
{  The effect of temperature on the electrical activity of the giant axon of
  the squid.}
J. of Physiology {\bf 109} 240-249 (1949)

\bibitem{FENG1936}
T.P. Feng:
{  The heat production of nerve.}
Ergebnisse der Physiologie, biologischen Chemie und experimentellen
Pharmakologie {\bf 38}, 73-132 (1936)

\bibitem{HOWARTH1968}
J.V. Howarth, R.D. Keynes and J.M. Ritchie:
{  The origin of the initial heat associated with a single impulse in
  mammalian non-myelinated nerve fibres.}
J. of Physiology {\bf 194} 745-793 (1968)

\bibitem{HOWARTH1975}
J.V. Howarth, R.D. Keynes and J.M. Ritchie and A. vin Muralt:
{  The heat production associated with the passage of a single impulse in
  olfactory nerve fibres.}
J. of Physiology {\bf 249}, 349-368 (1975)

\bibitem{DE-LICHTERVELDE}
A.C.L. de Lichtervelde, J.P. de Souza and M.Z. Bazant:
{  Heat of nervous conduction: A thermodynamic framework.}
PRE {\bf 101} 022406-1-13 (2020)

\bibitem{TASAKI1968}
I. Tasaki, A. Watanabe, R. Sandlin and L. Carnay:
{  Changes in fluorescence, turbidity and birefringence associated with nerve
  excitation.}
Proc. Natl. Acad. Sci. (USA) {\bf 61}, 883-888 (1968)

\bibitem{HILL-BC}
B.C. Hill, E.D. Schubert, M.A. Nokes and R.P. Michelson:
{  Laser Interferometer Measurement of Changes in Crayfish Axon Diameter
  Concurrent with Action Potential.}
Science {\bf 196}, 426-428 (1977)

\bibitem{IWASA}
K. Iwasa and I. Tasaki:
{  Mechanical changes in squid giant axons associated with production of
  action potential.}
Biochem. and Biophys. Res. Commun {\bf 95} 1328-1331 (1980)

\bibitem{IWASA2}
  K. Iwasa, I. Tasaki and R.C. Gibbons:
  {  Swelling of nerve fibers associated with action potentials.}
  Science {\bf 210}, 338-339 (1980)

\bibitem{TASAKI1992}
I. Tasaki and P.M. Byrne:
{  Discontinuous Volume Transitions in Ionic Gels and Their
  Possible Involvement in the Nerve Excitation Process.}
Biopolymers {\bf 32}, 1019-1023 (1992)

\bibitem{TASAKI1999a}
I. Tasaki:
{  Rapid Structural Changes in Nerve Fibers and Cells associated with Their
  Excitation Processes.}
Jap. J. of Physiology {\bf 49} 125-138 (1999)

\bibitem{TASAKI1999b}
I. Tasaki:
{  Evidence for phase transition in nerve fibers, cells and synapses.}
Ferroelectrics {\bf 220}, 305-316 (1999)

\bibitem{JENSEN}
M.{\O} Jensen, V. Jogini, D.W. Borhani, A.E. Leffler, 
R.O. Dror and D.E. Shaw:
{  Mechanism of Voltage Gating in Potassium Channels.}
Science {\bf 336} (6078) 229-233 (2012)

\bibitem{CATERALL2000}
W.A. Catterall:
{  From Ionic Currents to Molecular Mechanisms: The Structure and Function
of Voltage-Gated Sodium Channels.}
Neuron {\bf 26}, 13-25 (2000)

\bibitem{MPsolitons}
T. Dauxois and M. Peyrard: {  Physics of Solitons.} Cambridge University
Press, 2006

\bibitem{GONZALEZ-PEREZ2014}
A. Gonzalez-Perez, L.D. Mosgaard, R. Budvytyte, S. Nissen and T. Heimburg:
{  Penetration of Action Potentials During Collision in the Median and
  Lateral Giant Axons of Invertebrates.}
Phys. Rev. X {\bf 4} 031047 (2014)

\bibitem{TASAKI1949}
I. Tasaki:
{  Collision of two nerve impulses in the nerve fibre.}
Biochim. Biophys. Acta {\bf 3}, 494-497 (1949)

\bibitem{ASLANIDI}
O.V. Aslanidi and O.A. Mornev:
{  Can colliding nerve pulses be reflected?}
JETP Letters {\bf 65}, 579-585 (1997)
(Pis’ma Zh. Éksp. Teor. Fiz. 65, No. 7, 553–558 10 April 1997)

\bibitem{XU-2013}
Ke Xu, Guisheng Zhong and Xiaowei Zhuang:
{  Actin, Spectrin, and Associated Proteins Form a Periodic
  Cytoskeletal Structure in Axons.}
Science {\bf 339}, 452-456 (2013)

\bibitem{KOTTHAUS} 
J.P. Kotthaus: {  A Mechatronics view at nerve conduction.}
  arXiv:1909.06313 [physics.bio-ph] (2019)

\bibitem{PURCELL}
E.M. Purcell:
{  Life at low Reynolds number.}
Am. J. Phys. {\bf 45} 3-11 (1977)

\bibitem{NEHER}
E. Neher and B. Sakmann:
{  Single-channel currents recorded from
membrane of denervated frog muscle fibres.}
Nature {\bf 260}, 799-802 (1976)

\bibitem{EL-HADY}
A. El Hady and B.B. Machta:
{  Mechanical surface waves accompany action potential propagation.}
Nature Communications {\bf 6}:6697 (2015)

\bibitem{ENGELBRECHT}
J. Engelbrecht, T. Peets and K. Tamm:
{  Electromechanical coupling of waves in nerve fibres.}
Biomech. Model Mechanobiol. {\bf 17}, 1771-1783 (2018) and arXiv:1802.07014v2

\bibitem{FITZHUGH-1961}
R. FitzHugh:
{  Impulses and Physiological States in Theoretical Models of the Nerve
  Membrane.} 
Biophys. J. {\bf 1}, 445-466 (1961)

\bibitem{NAGUMO}
J. Nagumo, S. Arimoto and S. Yoshizawa:
{  An active pulse transmission line simulating nerve axon.}
Proceedings of the IRE, 2061-2070 (1962)

\bibitem{KRICHEN}
S. Krichen and P. Sharma:
{Flexoelectricity: A Perspective on an Unusual Electromechanical Coupling.} 
J. Appl. Mech. 83, 030801-1-6 (2016)

\bibitem{CHEN-Haoyu}
H. Chen, D. Garcia-Gonzalez, and A. J\'erusalem:
{Computational model of the mechanoelectrophysiological coupling in axons
  with application to neuromodulation.}
Phys. Rev. E {\bf 99} 032406 (18pp) (2019)

\bibitem{FRANKS1994}
N.P. Franks and W.R. Lieb:
{Molecular and cellular mechanisms of general anaesthesia.}
Nature {\bf 367} 607-614 (1994)

\bibitem{YAKAMURA}
T. Yakamura, E. Bertaccini, J.R. Trudell and R.A. Harris:
{Anesthetics and Ion Channels: Molecular Models and Sites of Action.}
Annual Review of Pharmacology and Toxicology {\bf 43}, 23-51 (2001)

\bibitem{EL-DIN}
T.M.G. El-Din, M.J.Lanaeus, N. Zheng and W.A. Catterall:
{Fenestrations control resting-state block of a voltage-
  gated sodium channel.}
PNAS {\bf 51} 13111-13116 (2018)

\bibitem{PAVEL}
M.A. Pavel, E.N. Petersen, H. Wang, R.A. Lerner and S.B. Hansen:
{Studies on the mechanism of general anesthesia.}
PNAS {\bf 117}, 13757-13766 (2020)

\bibitem{HEIMBURG2020} 
  T. Heimburg:
{The important consequences of the reversible heat production in
nerves and the adiabaticity of the action potential.}
arXiv:2002.06031v2 [physics.bio-ph] (2020)

\bibitem{BEYDER}
A. Beyder, J.L. Rae, C. Bernard, P.R. Strege,
F. Sachs and G. Farrugia:
{  Mechanosensitivity of Na v 1.5, a voltage-sensitive
  sodium channel.}
J. Physiol. {\bf 588} 4969-4985 (2010)

\bibitem{FITZHUGH-1962}
R. FitzHugh:
{  Computation of impulse initiation and saltatory conduction in a myelinated
nerve fiber.}
Biophys. J. {\bf 2}, 11-21 (1962)

\bibitem{ORI2018}
H. Ori, E. Marder and S. Marom:
{  Cellular function given parametric variation in the
  Hodgkin and Huxley model of excitability.}
PNAS {\bf 115}, E8211–E8218 (2018)

\bibitem{STRASSBERG}
A.E. Strassberg and L.J. DeFelice:
{  Limitations of the Hodgkin-Huxley Formalism: Effects of
Single Channel Kinetics on Transmembrane Voltage
Dynamics.}
Neural Computation {\bf 5}, 843-855 (1993)

\bibitem{MEUNIER}
C. Meunier and I. Segev:
{  Playing the Devil’s advocate: is the
  Hodgkin–Huxley model useful?}
TRENDS in Neurosciences {\bf 25}, 558-563 (2002)

\end{thebibliography}
\end{document}